\documentclass[11pt]{article}
\usepackage{amsmath,amsthm,amssymb,cite,enumerate}
\usepackage[a4paper,left=0.5in,right=1in]{geometry}
\usepackage[usenames]{color}
\usepackage{graphicx}
\usepackage{epsfig}
\usepackage{dcolumn}
\usepackage{bm}
\usepackage{authblk} 
\usepackage{bbold} 

\begin{document}

\def \d {{\rm d}}

\def \bF {\mbox{\boldmath{$F$}}}
\def \bH {\mbox{\boldmath{$H$}}}
\def \bV {\mbox{\boldmath{$V$}}}
\def \bff {\mbox{\boldmath{$f$}}}
\def \bT {\mbox{\boldmath{$T$}}}
\def \bk {\mbox{\boldmath{$k$}}}
\def \bl {\mbox{\boldmath{$\ell$}}}
\def \bn {\mbox{\boldmath{$n$}}}
\def \bbm {\mbox{\boldmath{$m$}}}
\def \tbbm {\mbox{\boldmath{$\bar m$}}}

\def \T {\bigtriangleup}
\newcommand{\msub}[2]{m^{(#1)}_{#2}}
\newcommand{\msup}[2]{m_{(#1)}^{#2}}

\newcommand{\be}{\begin{equation}}
\newcommand{\ee}{\end{equation}}

\newcommand{\beq}{\begin{eqnarray}}
\newcommand{\eeq}{\end{eqnarray}}
\newcommand{\pa}{\partial}
\newcommand{\pp}{{\it pp\,}-}
\newcommand{\ba}{\begin{array}}
\newcommand{\ea}{\end{array}}

\newcommand{\M}[3] {{\stackrel{#1}{M}}_{{#2}{#3}}}
\newcommand{\m}[3] {{\stackrel{\hspace{.3cm}#1}{m}}_{\!{#2}{#3}}\,}

\newcommand{\tr}{\textcolor{red}}
\newcommand{\tb}{\textcolor{blue}}
\newcommand{\tg}{\textcolor{green}}

\newcommand*\bg{\ensuremath{\boldsymbol{g}}}
\newcommand*\bE{\ensuremath{\boldsymbol{E}}}
\newcommand*\bh{\ensuremath{\boldsymbol{h}}}
\newcommand*\bR{\ensuremath{\boldsymbol{R}}}
\newcommand*\bu{\ensuremath{\boldsymbol{u}}}
\newcommand*\bA{\ensuremath{\boldsymbol{A}}}
\newcommand*\bep{\ensuremath{\boldsymbol{\varepsilon}}}

\def\a{\alpha}
\def\b{\beta}
\def\g{\gamma}
\def\de{\delta}

\def\E{{\cal E}}
\def\B{{\cal B}}
\def\R{{\cal R}}
\def\F{{\cal F}}
\def\L{{\cal L}}

\def\e{e}
\def\bb{b}

\newtheorem{theorem}{Theorem}[section] 
\newtheorem{cor}[theorem]{Corollary} 
\newtheorem{lemma}[theorem]{Lemma} 
\newtheorem{prop}[theorem]{Proposition}
\newtheorem{definition}[theorem]{Definition}
\newtheorem{remark}[theorem]{Remark}  
\newtheorem{proposition}[theorem]{Proposition}

\title{Universal $p$-form black holes in generalized theories of gravity}

\author[1]{Sigbj\o rn Hervik\thanks{sigbjorn.hervik@uis.no}}
\author[2]{Marcello Ortaggio\thanks{ortaggio(at)math(dot)cas(dot)cz}}

\affil[1]{Faculty of Science and Technology, University of Stavanger, N-4036 Stavanger, Norway}
\affil[2]{Institute of Mathematics of the Czech Academy of Sciences, \newline \v Zitn\' a 25, 115 67 Prague 1, Czech Republic}

\maketitle

\abstract{
We explore how far one can go in constructing $d$-dimensional static black holes coupled to $p$-form and scalar fields before actually specifying the gravity and electrodynamics theory one wants to solve. At the same time, we study to what extent one can enlarge the space of black hole solutions by allowing for horizon geometries more general than spaces of constant curvature. We prove that a generalized Schwarzschild-like ansatz  with an arbitrary isotropy-irreducible homogeneous base space (IHS)
provides an answer to both questions, up to naturally adapting the gauge fields to the spacetime geometry. In particular, an IHS-K\"ahler base space enables one to construct magnetic and dyonic 2-form solutions in a large class of theories, including non-minimally couplings. We exemplify our results by constructing simple solutions to particular theories such as $R^2$, Gauss-Bonnet and (a sector of) Einstein-Horndeski gravity coupled to certain $p$-form and conformally invariant electrodynamics. 
}

\vspace{.2cm}
\noindent





\section{Introduction}

\label{intro}

There has been a growing interest over the past two decades in black holes in more than four dimensions \cite{EmpRea08}. Moreover, effective descriptions of quantum corrections \cite{DeWittbook} and low-energy limits of string theory \cite{SchSch74} typically result in higher order terms to be added to the Einstein-Hilbert action, so that various gravity theories beyond general relativity have also been the subject of increasing attention.

In particular, when searching for static, spherically symmetric black holes in $d=n+2$ dimensions, one can start from a Schwarzschild-like ansatz of the form \cite{Tangherlini63,KraPle80} 
\be
 \d s^2=-A^2(r)\d t^2+B^2(r)\d r^2+r^2 h_{ij}(x)\d x^i\d x^j ,
 \label{metric}
\ee
i.e., a warped product $M=M_2\times M_n$, where the $n$-dimensional space $M_n$ carries a Riemannian metric $\bh=h_{ij}(x)\d x^i\d x^j$ (here $i,j,\ldots=1,2,\ldots,n$, and $x$ denotes collectively the set of coordinates $(x^1,\ldots,x^n)$ of $M_n$). Spherical symmetry requires $(M_n,\bh)$ to be a round sphere \cite{Tangherlini63,KraPle80}, however, metrics~\eqref{metric} with a different base space are also of interest (e.g., topological black holes).

In higher-dimensional Einstein gravity, the spherically symmetric Schwarzschild-Tangherlini solution \cite{Tangherlini63} has been generalized such that $\bh$ can in fact be {\em any} $n$-dimensional Einstein space \cite{GibWil87,Birmingham99}, which for $d>5$ gives rise to a much richer family of static ``Einstein'' black holes (at the price of giving up asymptotic flatness or (A)dS-ness). The property of $\bh$ being Einstein is not only sufficient but also necessary in order to satisfy Einstein's equations  (together with fixing the precise form of $B^{-2}=A^2=K-\mu r^{3-d}-\lambda r^2$), so that the extensions obtained in \cite{GibWil87,Birmingham99} in fact exhaust the space of black hole solution of the form~\eqref{metric} in general relativity. The particular choice of $\bh$ may affect the stability of the solution \cite{GibHar02}.

Similarly, it would be desirable to characterize the full space of solutions~\eqref{metric} also for more general theories of gravity -- ideally for any diffeomorphism invariant, metric theory. However, simply adding a Gauss-Bonnet \cite{DotGle05} or Lovelock \cite{FarDeh14,Ray15,OhaNoz15} term to the Einstein-Hilbert action places a stringent tensorial constraint on the geometry of $\bh$. This rules out various known ``exotic'' Einstein black holes and shows that, generically, $\bh$ cannot be an arbitrary Einstein space in a gravity theory different from Einstein's. 

Nevertheless, we have recently proved \cite{HerOrt20} (see also \cite{HerOrt20_proc} for a short summary) that there does exist a special family of Einstein metrics $\bh$ -- the {\em isotropy-irreducible homogeneous spaces} (IHS) \cite{Manturov61,Manturov61_2,Wolf68,Kramer75,WanZil91} (see section~\ref{sec_metric} for a definition) -- which is virtually immune to all possible tensorial constraints placed by any vacuum theory of gravity for which the Lagrangian is a scalar invariant constructed from the Riemann tensor $\bR$ and its covariant derivatives of arbitrary order. Namely, we have shown that in any such theory, for {\em any} IHS $\bh$ the ansatz \eqref{metric} ensures that the corresponding field equations automatically reduce to just two ODEs for the two unknown metric functions $A(r)$ and $B(r)$. This dramatically enlarges the space of vacuum black hole solutions and permitted horizon geometries for generalized theories of gravity, well beyond the usual case of horizons of constant curvature. We dubbed these spacetime {\em universal black holes} \cite{HerOrt20} to stress the theory-independence of this result.\footnote{The details (including the precise form of $A(r)$ and $B(r)$) and physical properties of the solutions will naturally depend on the specific theory one is interested in -- see \cite{HerOrt20} and references therein for various examples.}

Black holes with non-trivial gauge fields play an important role in supergravity and string theories (see, e.g., \cite{EmpRea08} and references therein). Along with higher-order gravity corrections, there are thus compelling reasons to consider theories containing also modifications of the Maxwell term in the action, together with a scalar field and possibly non-minimal couplings \cite{SchSch74,Callanetal85,FraTse85}. In general, these will be comprised in an action of the type
\be
 S=\int\d^dx\sqrt{-g}{\cal L}(\bR,\nabla\bR,\ldots,\bF,\nabla\bF,\ldots,\varphi,\nabla\varphi,\ldots) ,
\label{action}
\ee
where $\bF=\d\bA$ is a $p$-form and $\varphi$ a scalar field, and the Lagrangian density ${\cal L}$ is a scalar invariant. Even without introducing higher-order corrections, already in the Einstein-Maxwell theory the permitted horizon geometries for black holes~\eqref{metric} turn out to be constrained (in addition to being Einstein) already by the coupling to magnetic 2-form \cite{OrtPodZof08} or to electric and magnetic $p$-form fields \cite{BarCalCha12} (see also \cite{OrtPodZof15}).\footnote{The dual cases $p=1$ and $p=d-1$ are special and lead to static black holes with {\em non-Einstein} (and thus non-IHS) horizons \cite{OrtPodZof15}. These are not relevant to the discussion of the present paper, hence we hereafter assume $2\le p\le n$ (recall $d=n+2$).} More generally, a question arises as to whether a theory-independent characterization of 
the space of black hole solutions~\eqref{metric} in any theory of the form~\eqref{action} can be given (which would be a natural starting point for obtaining more refined horizon characterizations for specific theories). In this paper we present new results in this direction. In particular, we will show that the metric ansatz studied in \cite{HerOrt20} -- i.e., \eqref{metric} with $\bh$ being an IHS -- can also be consistently employed to construct $d$-dimensional static vacuum black hole solutions in any theory~\eqref{action}, provided the gauge fields are suitably adapted to the geometry.\footnote{A word of caution is in order here. Namely, properties of characteristic surfaces, and thus of light propagation and causality (in particular, event horizons) generally depend on the particular theory of electromagnetism one wants to consider -- see \cite{Boillat70,Plebanski70} for early results in the context of non-linear electrodynamics. Nevertheless, some universality properties of black hole event horizons have been pointed out in \cite{Shore95,GibHer01_prd}, at least for certain theories.} Apart from recovering in a unified way results previously obtained on a case-by-case basis for particular theories (such as Gauss-Bonnet and Lovelock gravity), this can thus be used, for example, to extend the well-known electrically charged higher-dimensional Reissner-Nordstr\"om-(A)dS spacetimes of general relativity \cite{Tangherlini63,GibWil87} and their magnetic counterparts \cite{OrtPodZof08} to any theory~\eqref{action}, as we will discuss.

The plan of the paper is as follows. In section~\ref{sec_metric} we first provide a definition of IHSs and summarize their basic properties (these are known facts but may be useful for readers not familiar with the topic). Then we discuss the form of the field equations for a theory of the type~\eqref{action} with an ansatz~\eqref{metric} (where $\bh$ is an IHS), and the main geometric properties of such spacetimes, to be employed in the following. In section~\ref{sec_magn_volume} we show that this geometry can support a magnetic $n$-form living on the base space, in which case the corresponding field equations reduce to two ODEs for the metric functions $A(r)$ and $B(r)$ (or three if a scalar field is also present). A similar discussion is presented in section~\ref{sec_electric} for a Coloumb-like radial electric 2-form -- one difference here being the presence of an additional ODE for the electric field. In section~\ref{sec_Kahl} we show that one can further extend the previous construction to magnetic and dyonic 2-forms if $\bh$ is additionally assumed to be K\"ahler, as well as to certain forms of higher rank in even dimensions. For the sake of definiteness, in section~\ref{sec_examples} we construct a few explicit solutions of gravity theories such as $R^2$, Gauss-Bonnet and a particular sector of Einstein-Horndeski gravity coupled to certain linear and non-linear electrodynamics. We conclude with a short summary and some additional comments in the last section~\ref{sec_discuss}.

\section{The metric ansatz}

\label{sec_metric}

Here we will discuss the form of the the field equations for a theory of the type~\eqref{action} for the metric ansatz~\eqref{metric}, where $\bh$ is assumed to be an IHS.\footnote{Note that generically one has to retain two independent functions $A(r)$ and $B(r)$ in the metric~\eqref{metric}. The special condition $A(r)B(r)=$const possesses an invariant characterization in terms of null alignment properties of the spacetime curvature \cite{HerOrt20,PodOrt06,PodSva15} and is thus non-generic. However, it is known to be satisfied in particular cases such as Einstein and (generic) Gauss-Bonnet and Lovelock gravities coupled to Maxwell fields. A class of four dimensional non-minimally coupled theories also admitting such kind of solutions has been recently proposed in \cite{CanMur20}.} Before proceeding, it may be useful to recall that an IHS is defined as a homogeneous connected Riemannian space $(M_n,\bh)$ such that, for every point $x\in M_n$, the isotropy group at $x$ (i.e., the isometries of $(M_n,\bh)$ leaving $x$ fixed) acts irreducibly on the tangent space of $M_n$ at $x$. More precisely, the space is {\em strongly} IHS  if also the identity component of the isotropy group acts irreducibly, and {\em weakly} IHS otherwise (the latter also include products of identical irreducible IHS). These have been classified, respectively, in \cite{Manturov61,Manturov61_2,Wolf68,Kramer75} and in \cite{WanZil91} (see also \cite{Bessebook}). All IHS are Einstein \cite{Wolf68}. Let us further mention that irreducible symmetric spaces (classified by Cartan, see, e.g., \cite{Wolf67,Bessebook}) are necessarily strongly IHS \cite{Manturov61,Manturov61_2,Wolf68,Kramer75}. Note also that $(M_n,\bh)$, being homogeneous, can only possess constant scalar invariants, and is necessarily compact if $K>0$ and necessarily non-compact if $K<0$ \cite{Bessebook}, and locally flat iff $K=0$. In addition, a non-compact IHS must be symmetric (with $K<0$) or flat \cite{Bessebook}. The simplest examples are the well-known spaces of constant curvature, or direct products of identical copies of those. The minimal dimension necessary for an IHS to be not of constant curvature is $n=4$, in which case an IHS must, however, be symmetric and therefore locally one of the following: $S^4$, $S^2\times S^2$, $H^4$, $H^2\times H^2$, $\mathbb{C}P^2$, $H_{\mathbb{C}}^2$, or flat space (cf. \cite{Bessebook} and references therein).

Let us now consider a generic theory of the type~\eqref{action}. Extremizing the action w.r.t. $\bg$, $\bA$ and $\varphi$, one obtains equations of motion of the form \cite{Eddington_book,Horndeski74_unp,Horndeski76,Anderson78,IyeWal94}
\beq
 & & \bE\equiv\frac{1}{\sqrt{-g}}\frac{\delta\left(\sqrt{-g}\cal L\right)}{\delta\bg}=0 , \label{grav_eq} \\
 & & \mbox{div}\bH\equiv\frac{\delta\cal L}{\delta\bA}=0 , \label{form_eq} \\
 & & \psi\equiv\frac{\delta\cal L}{\delta\varphi}=0 , \label{scal_eq}
\eeq
where $\bE$, $\bH$ and $\psi$ are, respectively, a symmetric 2-tensor, a $p$-form (with $2\le p\le n$) and a scalar field locally constructed in terms of (contractions of) $\bR$, $\bF$, $\varphi$ and their covariant derivatives of arbitrary order. When \eqref{form_eq} and \eqref{scal_eq} are satisfied, $\bE$ is conserved, i.e., $\nabla_\nu E^{\mu\nu}=0$ \cite{Eddington_book,Horndeski74_unp,Horndeski76,Anderson78,IyeWal94}.\footnote{The explicit form of the field equations \eqref{grav_eq}--\eqref{scal_eq}, which is determined after specifying the Lagrangian density ${\cal L}$ in \eqref{action}, is not needed for the purposes of the present paper. However, for the sake of definiteness, one simple example is provided by the $d=4$ (second-order) theory with a non-minimal coupling proposed by Horndeski \cite{Horndeski76}, i.e., ${\cal L}=\sqrt{-g}\left[\frac{1}{\kappa}(R-2\Lambda)-\beta F_{\mu\nu}F^{\mu\nu}-\gamma F_{\mu\nu}F^{\alpha\beta}\,{}^{*}\!R^{*\mu\nu}_{\phantom{*\mu\nu}\alpha\beta}\right]$, for which $E_{\mu\nu}=\frac{1}{\kappa}\left(R_{\mu\nu}-\frac{1}{2}Rg_{\mu\nu}+\Lambda g_{\mu\nu}\right)-2\beta\left(F_{\mu\rho}F_\nu^{\phantom{\nu}\rho}-\frac{1}{4}g_{\mu\nu}F_{\alpha\beta}F^{\alpha\beta}\right)-2\gamma\left(F^{\alpha\rho}F^{\beta}_{\phantom{\beta}\rho}\,{}^{*}\!R^*_{\mu\alpha\nu\beta}+{}^{*}\!F_{\mu\alpha;\beta}\,{}^{*}\!F_{\nu}^{\phantom{\nu}\beta;\alpha}\right)$ and $H^{\mu\nu}_{\phantom{\mu\nu};\nu}=4\beta F^{\mu\nu}_{\phantom{\mu\nu};\nu}+4\gamma F_{\beta\gamma;\alpha}\,{}^{*}\!R^{*\mu\alpha\beta\gamma}$ (here an asterisk denotes Hodge duality and $\kappa$, $\beta$ and $\gamma$ are coupling constants). When $\gamma=0$, this reduces to the standard Einstein-Maxwell theory. Another explicit example will be provided in section~\ref{subsec_GB}.}

Not surprisingly, the above equations simplify considerably for the ansatz~\eqref{metric}. Let us follow an approach similar to the one of \cite{Gurses92,GurSer95} (where $\bh$ was assumed to be a round sphere). If one defines the two covectors
\be
 \bu=\d t , \qquad \bk=\d r , 
\ee
the Riemann tensor $\bR$ of the spacetime~\eqref{metric} can be written as \cite{GurSer95} 
\beq
 R_{\mu\nu\rho\sigma}= \left(AA''-\frac{AA'B'}{B}\right)4u_{[\mu}k_{\nu]}u_{[\rho}k_{\sigma]}-\frac{4rAA'}{B^2}u_{[\mu}h_{\nu][\rho}u_{\sigma]}-\frac{4rB'}{B}k_{[\mu}h_{\nu][\rho}k_{\sigma]}  \nonumber \\
  {}-\frac{2r^2}{B^2}h_{\mu[\rho}h_{\sigma]\nu}+r^2\tilde R_{\mu\nu\rho\sigma} ,
 \label{riem}
\eeq
where, from now on, quantities with a tilde will refer to the transverse space geometry of $\bh$, and a prime denotes differentiation w.r.t. $r$.

Clearly $\bh$ has been promoted to a full spacetime tensor defined by
\be
 h_{\mu \nu}=r^{-2}\left(g_{\mu \nu}+A^2u_\mu u_\nu-B^2k_\mu k_\nu\right) , 
\ee
which implies $h_{\mu \nu}u^\nu=0=h_{\mu \nu}k^\nu$ (indices are raised with $g^{\mu\nu}$, therefore $u_\mu u^\mu=-A^{-2}$, $k_\mu k^\mu=B^{-2}$, $u_\mu k^\mu=0$). The covariant derivatives of $\bu$, $\bk$ and $\bh$ then read
\be
 \nabla_\nu u_\mu=-\frac{A'}{A}2u_{(\mu}k_{\nu)} , \qquad \nabla_\nu k_\mu=-\frac{1}{B^2}(BB'k_\mu k_\nu+AA'u_\mu u_\nu-rh_{\mu \nu}) , \qquad \nabla_\rho h_{\mu \nu}=-\frac{1}{r}h_{\mu \nu}k_\rho .
 \label{der_ukh}
\ee

Since these covariant derivatives can be fully expressed in terms of $\bu$, $\bk$ and $\bh$, the same will be true for covariant derivatives of any order. It is thus clear that any tensor constructed from $\bR$ (eq.~\eqref{riem}) and its covariant derivatives can be expressed in terms of $\bu$, $\bk$ and $\bh$ and covariant derivatives $\tilde\nabla^{(k)}\tilde\bR$ intrinsic to $\bh$. In particular, thanks to the time-reversal invariance of $\bg$, any such tensor can contain only an even number of $\bu$ (more generally, it must share the symmetries of $\bh$).

So far we have not used the fact that $\bh$ is IHS. The latter property further ensures that one cannot construct a non-zero vector out of $\tilde\nabla^{(k)}\tilde\bR$ and that the only possible symmetric 2-tensor that one can construct is $\bh$ itself (up to a constant factor) \cite{Wolf68}. This property will be crucial in what follows. Throughout the paper, we will also assume that the scalar field $\varphi=\varphi(r)$ depends only on the radial coordinate. Therefore,
\be
 \nabla_\tau\varphi=\varphi'k_\tau , 
\ee 
and (recalling also \eqref{der_ukh}) all higher-order covariant derivatives $\nabla^{(k)}\varphi$ can thus be expressed only in terms of $\bk$, $\bh$ and an even number of $\bu$.

\section{Magnetic fields of rank $p=n$}

\label{sec_magn_volume}

Let us consider here the case when the rank of $\bF$ in \eqref{action} is $p=n$ and assume that $\bF$ is purely magnetic and of the form
\be
 \bF=q\tilde\bep ,
 \label{F_magn}
\ee
where $q$ is a constant and $\tilde\bep$ is the volume element associated with $\bh$.\footnote{A more general ansatz with $q=q(r)$ is ruled out by $\bF=\d\bA$.} By construction $u^\mu F_{\mu\nu\ldots}=0=k^\mu F_{\mu\nu\ldots}$.

In this case the covariant derivative of $\bF$ reads 
\be
 \nabla_\tau F_{\mu\nu\ldots}=-\frac{n}{r}(k_\tau F_{\mu\nu\ldots}+F_{\tau[\nu\ldots}k_{\mu]}) .
\ee 
Recalling also \eqref{der_ukh}, one can argue that, similarly as in the case of the Riemann tensor, all higher-order covariant derivatives $\nabla^{(k)}\bF$ can be expressed only in terms of $\bF$, $\bu$, $\bk$ and $\bh$, and can contain only an even number of $\bu$.

Using also the fact that $(M_n,\bh)$ is IHS (see the comments in section~\ref{sec_metric} and \cite{HerOrt20} for a related discussion), it follows that any symmetric 2-tensor $\bE$ constructed from tensor products, sums and contractions from $\nabla^{(k)}\bR$, $\nabla^{(l)}\bF$ and $\nabla^{(m)}\varphi$ ($k,l,m\ge0$) must be of the form
\be  
	\bE=F(r)\bu\bu+G(r)\bk\bk+L(r)\bh . 
	\label{E}
\ee

Similarly, one can argue that the only $n$-form  $\bH$ that one can construct (automatically invariant under the isometries of $\bh$) can be written as
\be
 \bH=N(r)\bF .
\label{H_magn}
\ee
Since $\bF$ is divergencefree by construction~\eqref{F_magn}, thus is also $\bH$, and the generalized Maxwell equation~\eqref{form_eq} is automatically satisfied.

One thus only needs to solve the scalar equation \eqref{scal_eq} and the tensorial field equation \eqref{grav_eq}. In view of \eqref{E}, the latter reduces to three ``scalar'' equations $F(r)=0$, $G(r)=0$ and $L(r)=0$. However, similarly as in \cite{HerOrt20}, it is easy to see that the identity $\nabla_\nu E^{\mu\nu}=0$     implies that $L(r)=0$ holds automatically once $F(r)=0=G(r)$ and \eqref{scal_eq} are satisfied. We are thus left with just {\em three ODEs for the two metric functions $A(r)$ and $B(r)$ and for the scalar field $\varphi(r)$}. Their precise form will depend on the particular theory~\eqref{action} one wants to study, and is of no interest for the general considerations of this paper. In the special case when $\bh$ is a round sphere, a similar result was obtained in \cite{Gurses92,GurSer95}.

Solutions of this type include, for instance, the magnetic duals \cite{BarCalCha12,OrtPodZof15} of electrically charged black holes of \cite{Tangherlini63,GibWil87} in the Einstein-Maxwell theory, and of similar solutions in Gauss-Bonnet-Maxwell gravity \cite{Wiltshire86,MaeHasMar10,BarChaKol11,CveNojOdi02} -- cf. also section~\ref{sec_electric}.

\section{Electric fields of rank $p=2$}

\label{sec_electric}

Here we consider the following purely electric 2-from
\be
 \bF=M(r)\bu\wedge\bk ,
 \label{F_elect}
\ee
such that $u^\mu F_{\mu\nu}=-A^{-2}Mk_\nu$ and $k^\mu F_{\mu\nu}=-B^{-2}Mu_\nu$.

One finds 
\be
 \nabla_\tau F_{\mu\nu}=\left[\frac{M'}{M}-\frac{(AB)'}{AB}\right]F_{\mu\nu}k_\tau+\frac{rM}{B^2}2u_{[\mu}h_{\nu]\tau} ,
\ee 
and all $\nabla^{(k)}\bF$ thus contain an odd number of $\bu$. 

From now on, when an electric field is present, we restrict ourselves to theories whose Lagrangian contains {\em only even powers of $\bF$ and its covariant derivatives}.\footnote{Apart from the standard Maxwell theory and its $p$-form extensions, this includes, for example, the power-like non-linear electrodynamics considered in \cite{CveNojOdi02,HasMar07,MaeHasMar10} and models often employed in string theory such as the Born-Infeld action (cf., e.g., \cite{Tseytlin00} and references therein).} One can then argue that the $2$-form $\bH$ (defined by \eqref{form_eq} extremizing~\eqref{action} w.r.t. $\bA$) must be purely electric (and invariant under the isometries of $\bh$) and thus of the form
\be
 \bH=N(r)\bF .
\ee
The $r$-component of \eqref{form_eq} is therefore automatically satisfied, while the $t$-components implies
\be
 M(r)N(r)=-\frac{eAB}{r^n} ,
 \label{eq_electr}
\ee
where $e$ is a constant. This is one of the field equations that has to be solved (the function $N$ contains in general $\bg$, $\bA$  and $\varphi$ and their derivatives, and its precise form depends on the particular theory one is considering). 

The remaining field equations come from the generalized Einstein equation~\eqref{grav_eq}. The same argument used in section~\ref{sec_magn_volume} enables one to  arrive again at~\eqref{E} (this would not be true if the Lagrangian contained also odd powers of $\bF$, which may result in $\bE$ containing an additonal term proportional to $\bu\bk+\bk\bu$, cf. also \cite{Gurses92} for related comments). Thanks to $\nabla_\nu E^{\mu\nu}=0$, we are finally left with {\em four ODEs for the two metric functions $A(r)$ and $B(r)$, for the scalar field $\varphi(r)$ and for the vector potential $\bA$} (i.e., $F(r)=0=G(r)$ and eqs.~\eqref{scal_eq} and \eqref{eq_electr}).

Note that when $N'=0$ and $AB=1$ (as in the Einstein-Maxwell theory), the field \eqref{F_elect} with \eqref{eq_electr} describes the standard Coulomb solution \cite{Tangherlini63,GibWil87}. Extensions are known in a number of generalized gravity theories -- for example, electric black holes in Gauss-Bonnet gravity in arbitrary dimensions with a spherical base space were obtained in \cite{Wiltshire86} (extended to zero and negative constant curvature in \cite{CveNojOdi02}), and with a more general base space in \cite{MaeHasMar10} (see also \cite{BarChaKol11} in six dimensions).

\section{IHS-K\"ahler base spaces: dyonic 2-form fields and forms of higher rank}

\label{sec_Kahl}

Let us consider the case when $n$ is {\em even} and the base space is both IHS and K\"ahler. Let us further assume that the K\"ahler 2-form $\mbox{\boldmath{$J$}}$ is invariant under the isometries of $\bh$.\footnote{In a compact (and orientable) space $M_n$, any harmonic form (thus the K\"ahler one in particular) is invariant under motions \cite{Yanobook}, therefore our assumption only possibly restricts the case when $M_n$ is non-compact (see \cite{Moroianubook} for an example violating this assumption).} Examples include the direct product of identical 2-spaces of constant curvature $S^2\times S^2\times\ldots$ and $H^2\times H^2\times\ldots$ (which are clearly reducible), or the complex projective space $\mathbb{C}P^{\frac{n}{2}}$ (compact) and the complex hyperbolic space $H_{\mathbb{C}}^{\frac{n}{2}}$ (non-compact) with the Fubini-Study metric \cite{KobNom2} (which are irreducible), and direct products of identical copies of those. These examples also happen to be symmetric spaces \cite{Wolf67,Bessebook} (see in particular \cite{Wolf67} for a discussion of irreducible K\"ahler symmetric spaces), while non-symmetric strongly IHS cannot occur here \cite{Wolf68}.

\subsection{Magnetic $p=2$ solutions}

\label{subsec_magn_p=2}

One can naturally define a magnetic 2-form field as a constant multiple of the K\"ahler form $\mbox{\boldmath{$J$}}$, i.e.,
\be
 \bF=q\mbox{\boldmath{$J$}} ,
 \label{F_Kahl}
\ee  
which is invariant under isometries by assumption. Thanks to the well-known properties (in $(M_n,\bh)$) $\mbox{\boldmath{$J^2$}}=-\mathbb{1}$ and $\tilde\nabla\mbox{\boldmath{$J$}}=0$, one gets easily
\be
 F_{\mu}^{\phantom\mu\rho}F_{\nu\rho}=q^2r^{-2}h_{\mu\nu} , \qquad \nabla_\tau F_{\mu\nu}=-\frac{2}{r}(k_\tau F_{\mu\nu}+F_{\tau[\nu}k_{\mu]}) .
 \label{FF_Kahl}
\ee
Similarly as in section~\ref{sec_magn_volume}, all higher-order covariant derivatives $\nabla^{(k)}\bF$ can thus be expressed only in terms of $\bF$, $\bu$, $\bk$ and $\bh$, and can contain only an even number of $\bu$. 

Therefore, any tensor constructed from  $\nabla^{(k)}\bR$, $\nabla^{(l)}\bF$ and $\nabla^{(m)}\varphi$ can  be expressed in terms of $\tilde\nabla^{(s)}\tilde\bR$, $\bF$,  $\bk$, $\bh$, and an even number of $\bu$, and is automatically invariant under isometries. 
Using a result of \cite{Wolf68} it thus follows again that any such symmetric 2-tensor can be written as in \eqref{E}. It also follows that, in $M_n$, $\widetilde{\mbox{div}}\bH=0$, otherwise this would define a preferred spatial vector in $M_n$, contradicting the IHS assumption. This suffices to show that the generalized Maxwell equation \eqref{form_eq} is identically satisfied (since $\bH$ lives in $M_n$). Summarizing, one is left with {\em three ODEs for $A(r)$, $B(r)$ and $\varphi(r)$} (i.e., $F(r)=0=G(r)$ and eq.~\eqref{scal_eq}).

For example, magnetic black hole solutions of this type were constructed in Einstein gravity in \cite{OrtPodZof08}, in Gauss-Bonnet gravity in \cite{MaeHasMar10} (including $F^4$ corrections) and in \cite{BarChaKol11}, and in Lovelock gravity in \cite{OhaNoz15}.

\subsection{Dyonic $p=2$ solutions}

Similarly as in the Einstein-Maxwell case \cite{OrtPodZof08}, one can superimpose the above magnetic field to the electric one of section~\ref{sec_electric}  to construct {\em dyonic} solutions of the type 
\be
 \bF=M(r)\bu\wedge\bk+q\mbox{\boldmath{$J$}} .
 \label{F_dyonic}
\ee

Here the most general $2$-form $\bH$ that one can construct is of the form
\be
 \bH=N(r)M(r)\bu\wedge\bk+\hat N(r)\mbox{\boldmath{$S$}} ,
\ee
where $\mbox{\boldmath{$S$}}$ is a 2-form in $M_n$ invariant under the isometries of $\bh$.
The field equations are the same as for the purely magnetic case~\eqref{F_Kahl} plus the additional ODE \eqref{eq_electr}. Recall, however, that when electric fields are present our results apply only to theories containing only even powers of $\bF$ (see section~\ref{sec_electric}). See again \cite{OrtPodZof08,MaeHasMar10,BarChaKol11,OhaNoz15} for examples in Einstein, Gauss-Bonnet and Lovelock gravity.

\subsection{Solutions with higher rank forms}

\label{subsec_higher}

Similarly as discussed in \cite{BarCalCha12} for the Einstein-Maxwell case, the above construction can be extended by considering the exterior product of $\mbox{\boldmath{$J$}}$ with itself $m$ times (with $2\le2m\le n$), giving rise to a magnetic $2m$-form
\be
 \bF=q\mbox{\boldmath{$J$}}\wedge\ldots\wedge\mbox{\boldmath{$J$}}.
 \label{F2m}
\ee  

The same argument used in section~\ref{subsec_magn_p=2} implies again that any symmetric 2-tensor that one can construct can be written as in \eqref{E}. To deal with the generalized Maxwell equation~\eqref{form_eq}, let us further assume here that $(M_n,\bh)$ is also {\em symmetric} (see \cite{Wolf67} for comments on irreducible K\"ahler symmetric spaces). Since any $\bH$,  by construction, will have only components in $M_n$ and will be invariant under the isometries of $\bh$, it follows \cite{KobNom1} that $\tilde\nabla\mbox{\boldmath{$H$}}=0$. This suffices to conclude that \eqref{form_eq} is identically satisfied.

We observe that other magnetic Ans\"atze for $\bF$ are also possible. For example, one can use $\star$-duality in $(M_n,\bh)$ to construct a $(d-4)$-form $\star$-dual to \eqref{F_Kahl}, or a $(d-2-2m)$-form $\star$-dual to \eqref{F2m}. The corresponding $\bH$ will automatically be purely magnetic and invariant under the isometries of $\bh$, and therefore again $\tilde\nabla\mbox{\boldmath{$H$}}=0$, so that the argument used above for \eqref{F2m} also applies here.

\section{Examples}

\label{sec_examples}

In this section we provide a few explicit examples of the general constructions described in the previous sections for some particular theories of gravity and electrodynamics (it is not our purpose here to discuss in detail the physical properties of such solutions -- related discussions for similar solutions of the same theories can be found in the references given in the following). 

In all examples the metric will be given by \eqref{metric}, where $\bh$ is IHS and normalized such that
 \be
 \tilde{R}_{ij}=(n-1)Kh_{ij} ,
 \label{def_K}
\ee
which gives $\tilde{R}=n(n-1)K$. Recall that $d=n+2$. Coupling constants of the various considered theories will be denoted by $\kappa$, $\alpha$, $\beta$, $\gamma$ and $\eta$, and the cosmological constant by $\Lambda$, while $\mu$ will denote an integration constant (in general related to the mass).

\subsection{Einstein-$R^2$ gravity with conformally invariant form fields}

Actions quadratic in the curvature have been studied for a long time \cite{Weyl18,Eddington_book,Lanczos38,Gregory47,Buchdahl48}. The simplest quadratic gravity theory comprises only the $R^2$ term, for which the field equations are of the fourth order but still relatively simple \cite{Gregory47,Buchdahl48}.

\subsubsection{Magnetic $2m$-form solution}

Let us consider Einstein gravity with an $R^2$ correction coupled to a $2m$-form in dimension $d=4m$, defined by the Lagrangian density
\be
  {\cal L}= \sqrt{-g} \left[\frac{1}{\kappa} \left( R - 2 \Lambda \right)+ \alpha R^2-\beta F_{\mu_1\ldots\mu_{2m}}F^{\mu_1\ldots\mu_{2m}}\right] .
	\label{R2_2mform}
\ee

It is not difficult to construct a magnetic solution with $A^2=B^{-2}$, $\bh$ taken to be IHS-K\"ahler (Einstein-K\"ahler actually suffices here), and $\bF$ given by \eqref{F2m}. The metric is then specified by 
\be
 A^2=B^{-2}=K-\lambda r^2-\frac{\mu}{r^{d-3}}+\frac{1}{r^{d-2}}\frac{\kappa\beta}{1+2d(d-1)\alpha\kappa\lambda}\frac{\hat q^2}{d-2}  , 
 \label{A_R2_2mform}
\ee
where for simplicity we have defined a rescaled parameter $\hat q$ such that $F_{\mu\nu\ldots}F^{\mu\nu\ldots}=\hat q^2r^{-d}$. The effective cosmological constant $\lambda$ is determined in terms of the coupling constants $\Lambda$, $\kappa$ and $\alpha$ by
\be
 2\Lambda=(d-1)\lambda\left[d-2+d(d-1)(d-4)\alpha\kappa\lambda\right] ,
 \label{lambda}
\ee
thus generically giving rise to two branches of solutions.\footnote{As seen from \eqref{A_R2_2mform}, one has to exclude the fine-tuned case $1+2d(d-1)\alpha\kappa\lambda=0$, which gives $1+8\kappa\alpha\Lambda=0$, for which the theory~\eqref{R2_2mform} admits special vacuum solutions (cf. \cite{Ayon-Beatoetal10,HerOrt20} and references therein).}  The simplicity of this solution is due to the fact that the energy-momentum of $\bF$ is traceless, which allows for spacetimes with constant Ricci scalar $R=d(d-1)\lambda$ -- the field equations then become effectively equivalent to those of the standard Einstein-Maxwell $p$-form theory \cite{BarCalCha12,OrtPodZof15} (up to a simple rescaling of $\kappa$ and $\Lambda$). It is thus easy to extend the above solution to dyonic fields of the form $\bF=M(r)\bu\wedge\bk\wedge\underbrace{\mbox{\boldmath{$J$}}\wedge\ldots\wedge\mbox{\boldmath{$J$}}}_{m-1}+q\underbrace{\mbox{\boldmath{$J$}}\wedge\ldots\wedge\mbox{\boldmath{$J$}}}_{m}$, similarly as in \cite{BarCalCha12}. An explicit example for the base space metric is given by a product of $(2m-1)$ identical 2-spheres of radius $a$, which can be written, e.g., as $\bh=\sum_{i=1}^{2m-1}\left[\left(1-\frac{\rho_i}{a^2}\right)^{-1}\d\rho_i^2+\left(1-\frac{\rho_i}{a^2}\right)\d\psi_i^2\right]$, so that $\mbox{\boldmath{$J$}}=\sum_{i=1}^{2m-1}\d\rho_i\wedge\d\psi_i$.

\subsubsection{Magnetic 2-form solution}

Similar ideas can be used to construct magnetic solutions of Einstein-$R^2$ gravity coupled to the conformally invariant non-linear electrodynamics of \cite{HasMar07} , i.e., 
\be
  {\cal L}= \sqrt{-g} \left[\frac{1}{\kappa} \left( R - 2 \Lambda \right)+ \alpha R^2-\beta(F_{\mu\nu}F^{\mu\nu})^{d/4}\right] .
	\label{R2_2form}
\ee

For this theory, the generalized Maxwell equations are given by \eqref{form_eq} with $\bH=d\beta(F_{\mu\nu}F^{\mu\nu})^{d/4-1}\bF$. Assuming $d$ to be even and $\bh$  Einstein-K\"ahler, $\bF$ is given by \eqref{F_Kahl} and
\be
 A^2=B^{-2}=K-\lambda r^2-\frac{\mu}{r^{d-3}}+\frac{1}{r^{d-2}}\frac{\kappa\beta}{1+2d(d-1)\alpha\kappa\lambda}\frac{\hat q^{d/2}}{d-2}  , 
	\label{A_R2_2form}
\ee
again with \eqref{lambda}, where we have normalized $F_{\mu\nu}F^{\mu\nu}=\hat q^2r^{-4}$. An electric solution can also be easily written down using the one given  for Einstein gravity in \cite{HasMar07} (in the purely electric case $\bh$ can be any Einstein space, not necessarily K\"ahler).

Both the above examples \eqref{A_R2_2mform} and \eqref{A_R2_2form} can be straightforwardly extended to more general $f(R)$ gravities, owing to $R$ being constant (see \cite{Sheykhi12} for electric solutions of the theory ${\cal L}= \sqrt{-g}\left[f(R)-\beta(F_{\mu\nu}F^{\mu\nu})^{d/4}\right]$ in the special case of spherical symmetry). Pure $R^2$-gravity solutions can be obtained easily taking the limit $\kappa\to\infty$ in \eqref{A_R2_2mform} or \eqref{A_R2_2form} (and suitably rescaling $\Lambda$ in \eqref{lambda}).

\subsection{Gauss-Bonnet gravity with a conformally invariant 2-form field}

\label{subsec_GB}

Apart from $R^2$ gravity, another interesting sector of quadratic gravity is given by Gauss-Bonnet gravity, which possesses field equations of the second order and is of particular interest in the low-energy limit of string theory \cite{Zwiebach85}. When it is coupled to the standard Maxwell theory, electrically charged and dyonic black holes have been studied in \cite{Wiltshire86,CveNojOdi02,MaeHasMar10,BarChaKol11}. Let us consider instead a coupling to the non-linear electrodynamics of \cite{HasMar07}, i.e., the theory
\be
 {\cal L}=\sqrt{-g}\left[\frac{1}{\kappa}(R-2\Lambda)+\gamma I_{GB}-\beta(F_{\mu\nu}F^{\mu\nu})^{d/4}\right] , 
 \label{GB}
\ee
where $I_{GB}=R_{\mu\nu\rho\sigma}R^{\mu\nu\rho\sigma}-4R_{\mu\nu}R^{\mu\nu}+R^2$. The field equations~\eqref{grav_eq} are determined by
\beq
 & & E_{\mu\nu}=\frac{1}{\kappa}\left(R_{\mu\nu}-\frac{1}{2}Rg_{\mu\nu}+\Lambda g_{\mu\nu}\right)+2\gamma\left(RR_{\mu\nu}-2R_{\mu\rho\nu\sigma}R^{\rho\sigma}+R_{\mu\rho\sigma\tau}R_\nu^{\ \rho\sigma\tau}-2R_{\mu\rho}R_\nu^{\ \rho}-\frac{1}{4}I_{GB}g_{\mu\nu}\right) \nonumber \\
 & & \qquad\qquad\qquad\qquad\qquad\qquad {}-2\beta\left(\frac{d}{4}F_{\mu\rho}F_\nu^{\phantom{\nu}\rho}(F_{\alpha\beta}F^{\alpha\beta})^{d/4-1}-\frac{1}{4}g_{\mu\nu}(F_{\alpha\beta}F^{\alpha\beta})^{d/4}\right) ,
 \label{E_GB}
\eeq
while in \eqref{form_eq} one should substitute $\bH=d\beta(F_{\mu\nu}F^{\mu\nu})^{d/4-1}\bF$.

Using metric~\eqref{metric} with the dyonic ansatz \eqref{F_dyonic}, eq.~\eqref{form_eq} reduces to \eqref{eq_electr}. Using \eqref{riem}, the $E^t_{\phantom{t}t}-E^r_{\phantom{r}r}$ component of \eqref{E_GB} gives simply (for a generic coupling constant $\gamma$)
\be
  (AB)'=0 , 
\ee
which without losing generality allows one to take $B=A^{-1}$ (as in the vacuum case \cite{DotGle05}). Using this, a solution to \eqref{eq_electr} and to the $E^t_{\phantom{t}t}$ component of \eqref{E_GB} is given by \eqref{F_dyonic} with
\beq
 & & M=-\frac{e}{r^2} , \\
 & & A^2=B^{-2}=K+\frac{r^2}{2\kappa\hat\gamma}\left[1\pm\sqrt{1+4\kappa\hat\gamma\left[\lambda+\frac{\mu}{r^{d-1}}-\frac{\kappa\hat\gamma\tilde I_W^2}{r^4}-\frac{\kappa\beta(\hat q^2-2e^2)^{(d-4)/4}}{r^d}\left(\frac{\hat q^2}{d-2}+e^2\right)\right]}\right] , \nonumber 
 \label{f_GB}
\eeq
where $\hat\gamma=(d-3)(d-4)\gamma$, $(d-1)(d-2)\lambda=2\Lambda$ and $(d-2)(d-3)(d-4)(d-5)\tilde I_W^2=\tilde C_{ijkl}\tilde C^{ijkl}$ (the constant $e$ differs by the same symbol in \eqref{eq_electr} by a numerical factor). The base space $\bh$ is even-dimensional and IHS-K\"ahler (it can be any IHS if $\hat q=0$). If $\hat q^2-2e^2<0$, $d$ must be a multiple of four, while for $\hat q^2-2e^2=0$ (i.e., $F_{\mu\nu}F^{\mu\nu}=0$) the electromagnetic field is stealth.\footnote{It is well-known that non-trivial stealth electromagnetic $p$-form fields are not possible in the standard Maxwell theory in any $d$-dimensional Lorentzian space (simply because $T_{\mu\nu}u^\mu u^\nu>0$ for any timelike $\bu$, unless $\bF=0$ \cite{Senovilla00}). See \cite{Smolic18} for a discussion of stealth fields in non-linear electrodynamics in four dimensions.} The branch with the minus sign in $A^2$ admits an Einstein gravity limit $\hat\gamma\to0$, which gives rise to a dyonic solution generalizing the electric black hole of \cite{HasMar07}. In the special case $\hat q=0=\tilde I_W^2$ one recovers a solution of \cite{Hendi09}. Magnetic solutions for other power-like electrodynamics have been obtained in \cite{MaeHasMar10}.

\subsection{Einstein-Horndeski theory}

Horndeski theory \cite{Horndeski74} is the most general scalar-tensor theory possessing field equations of the second order in four dimensions, and has received increasing attention in recent years. Particular sectors of it have been studied also in higher dimensions. A particularly simple one (see \cite{Sushkov09} and references therein), coupled to Einstein-Maxwell gravity, enables us to present explicit solutions in the presence of a 2-form and a non-minimally coupled scalar field, namely 
\be
  {\cal L}= \sqrt{-g} \left[\frac{1}{\kappa} \left( R - 2 \Lambda \right)+\eta G_{\mu\nu}\nabla^\mu\varphi\nabla^\nu\varphi-\beta F_{\mu\nu}F^{\mu\nu} \right] ,
	\label{Horn_Maxw}
\ee
where $G_{\mu\nu}$ is the Einstein tensor. 

Similarly as above, in even dimensions we take $\bh$ to be IHS-K\"ahler (Einstein-K\"ahler suffices) and such that $\tilde R\neq0$, and $\bF$ as in \eqref{F_Kahl} (with $F_{\mu\nu}F^{\mu\nu}=\hat q^2r^{-4}$). Then the metric and the scalar field are determined by  
\beq
  & & A^2=(d^2-1)\tilde R^2-2(d+1)(d-3)\Lambda\tilde R r^2+(d-1)(d-3)\Lambda^2r^4-\frac{\mu}{r^{d-3}} \nonumber \\
	& & \qquad\qquad {}+(d^2-1)\hat q^2\beta\kappa\left[\Lambda-\frac{d-3}{d-5}\frac{\tilde R}{r^2}+\frac{d-3}{4(d-7)}\frac{\beta\kappa\hat q^2}{r^4}\right] ,  \\
	& & B^{2}=(d-2)\frac{r(A^2)'+(d-3)A^2}{\tilde RA^2} , \qquad \varphi'^2=-\frac{2\Lambda r^4+\beta\kappa\hat q^2}{\eta\kappa\tilde R}\frac{B^2}{r^2}  . \nonumber 
\eeq
In four dimensions this reduces (up to electromagnetic duality) to a solution of \cite{CisEri14}. Electric solutions in higher dimensions with $\Lambda=0$ have been obtained in \cite{Fengetal16}, which by duality can be transformed into magnetic fields of the form \eqref{F_magn}.

A different behaviour is found if one considers again the non-linear electrodynamics \cite{HasMar07} mentioned above. Namely, for the theory 
\be
  {\cal L}= \sqrt{-g} \left[\frac{1}{\kappa} \left( R - 2 \Lambda \right)+\eta G_{\mu\nu}\nabla^\mu\varphi\nabla^\nu\varphi-\beta(F_{\mu\nu}F^{\mu\nu})^{d/4} \right] ,
	\label{Horn_HasMar}
\ee
one finds
\beq
  & & A^2=(d^2-1)\tilde R^2-2(d+1)(d-3)\Lambda\tilde R r^2+(d-1)(d-3)\Lambda^2r^4-\frac{\mu}{r^{d-3}} \nonumber \\
	& & \qquad\qquad {}+(d+1)(d-3)\frac{\hat q^{d/2}\beta\kappa}{r^{d-4}}\left[(d-1)\Lambda+(d-1)\frac{\tilde R}{r^2}-\frac{\beta\kappa}{4}\frac{\hat q^{d/2}}{r^d}\right] ,  \\
	& & B^{2}=(d-2)\frac{r(A^2)'+(d-3)A^2}{\tilde RA^2} , \qquad \varphi'^2=-\frac{2\Lambda r^d+\beta\kappa\hat q^{d/2}}{\eta\kappa\tilde R}\frac{B^2}{r^{d-2}}  \qquad (d=4m) . \nonumber 
\eeq
Electric solutions in higher dimensions have been obtained in \cite{Stetsko19} (also for more general powers of $F_{\mu\nu}F^{\mu\nu}$ in \eqref{Horn_HasMar}).

\section{Discussion}

\label{sec_discuss}

We have shown that one can go quite far in constructing $d$-dimensional static black holes coupled to $p$-form and scalar fields even before specifying the gravity and electrodynamics theory one wants to solve. Namely, a generalized Schwarzschild-like ansatz can be consistently employed to find $d$-dimensional static black hole solutions in {\em any} metric theory of gravity~\eqref{action} coupled to a $p$-form and a scalar field, up to adapting the matter fields to the spacetime geometry. This means that, irrespective of the theory one considers, the field equations reduce to four ODEs for two metric function, the scalar field and the electric part of the gauge field. Additionally, we have shown that this allows one to replace the standard spherical base space metric by an arbitrary isotropy-irreducible homogeneous space, giving rise to large families of static solutions and dramatically enlarging the space of permitted horizon geometries. Since we arrived at our conclusions without the need of specifying the explicit form of the underlying equations of motion, our results apply to general higher-derivative theories, for which constraints on the horizon geometry may generically contain an arbitrary number of covariant derivatives of the Riemann tensor. An extension to the case of multiple gauge and scalar fields can be worked out similarly. 

By extending our previous results for the vacuum case \cite{HerOrt20}, the present paper presents new results in the direction of providing a theory-independent characterization of permitted horizon geometries of static black holes. Given a specific theory, these results can be used as a starting point for obtaining a full horizon characterization thereof (this  may differ from our conclusions, since certain horizon geometries may be permitted in some theories but not in others, thus not being universal).

We have constructed examples of our results in a few theories of considerable interest, but the same methods can also be applied in other contexts. Just as an example, there is a growing interest also in (higher derivative) modifications of Lovelock's gravity such as quasi-topological gravities \cite{OlivaRay10,OlivaRay10_prd,MyeRob10,Dehghanietal12,Cisternaetal17,BueCanHen20}, to which our results also apply.

To conclude, it should be emphasized that physical properties of the black holes one can construct as described above, such as their thermodynamics and stability, will in general depend on the considered theory. Furthermore, while one has quite some freedom in choosing a IHS metric for the base space, this may affect the stability of the corresponding solution (see for example \cite{GibHar02} in Einstein gravity).

\section*{Acknowledgments}

We thank Gregory Horndeski for kindly making available to us his unpublished work \cite{Horndeski74_unp}. M.O. is also grateful to Sameer Murthy for a useful discussion. S.H. was supported through the Research Council of Norway, Toppforsk
grant no. 250367: \emph{Pseudo-Riemannian Geometry and Polynomial Curvature Invariants:
Classification, Characterisation and Applications.}  M.O. was supported by research plan RVO: 67985840 and research grant GA\v CR 19-09659S. 


\providecommand{\href}[2]{#2}\begingroup\raggedright\endgroup

\end{document}